\documentclass[twocolumn]{aastex631}

\newcommand{\msun}{{ M}_\odot}
\newcommand{\zsun}{{ Z}_\odot}
\usepackage{newtxtext,newtxmath}

\usepackage[normalem]{ulem}

\title{Metal-enriched Pair-instability supernovae: Effects of rotation}
\shortauthors{Nagele, Umeda and Maeda}
\graphicspath{{./}{}}

\begin{document}

\title{STELLA lightcurves of energetic pair instability supernovae in the context of SN2018ibb}

\author{Chris Nagele}
\email{chrisnagele.astro@gmail.com }
\affiliation{Department of Astronomy, School of Science, The University of Tokyo, 7-3-1 Hongo, Bunkyo, Tokyo 113-0033, Japan}
\author{Hideyuki Umeda}
\affiliation{Department of Astronomy, School of Science, The University of Tokyo, 7-3-1 Hongo, Bunkyo, Tokyo 113-0033, Japan}
\author{Keiichi Maeda}
\affiliation{Department of Astronomy, Kyoto University, Kitashirakawa-Oiwake-cho, Sakyo-ku, Kyoto 606-8502}



\begin{abstract}

SN2018ibb is a recently observed hydrogen poor super-luminous supernova which appears to be powered by the decay of $30\;\rm{M_\odot}$ of radioactive nickel. This supernova has been suggested to show hybrid signatures of a pair instability supernova and an interacting supernova. In a previous paper, we found that rotating, metal enriched pair instability supernova progenitors appeared to check both of these boxes. In this paper, we model the lightcurves of the pair instability supernovae using STELLA. We find that the STELLA models can explain the overall shape of the bolometric lightcurve of SN2018ibb, though not specific morphological features such as the luminosity peak or the bump at roughly three hundred days after the peak. We also estimate the contribution from interaction, and find that with relatively low wind velocities, the circum-stellar medium originating from the stellar winds is consistent with the evidence for interaction in the spectra. The observed values of the photosphere velocity in the hundred days after peak luminosity are similar to the STELLA models, but the deceleration is lower. This leads to the biggest inconsistency which is the black body temperature of SN2018ibb being much hotter than any of the STELLA models. We note that this high temperature (and the flat velocity) may be difficult to reconcile with the long rise time of SN2018ibb, but nevertheless conclude that if it is accurate, this discrepancy represents a challenge for SN2018ibb being a robust PISN candidate. This result is noteworthy given the lack of other scenarios for this supernova. 
 
\end{abstract}

\keywords{supernovae : general --- supernovae : individual : SN2018ibb --- 
  stars : abundances --- stars : rotation --- stars : black holes --- nucleosynthesis} 


\section{\textbf{Introduction}}
\label{introduction}

The lack thus far of clear evidence for pair instability supernovae (PISNe, \citealt{Barkat1967PhRvL..18..379B}) is puzzling. These hypothesized explosions are thought to occur when a massive carbon oxygen core (80-100 $\msun$) with consequent high entropy reaches a high enough temperature for electron positron pairs to be created \citep{Umeda2002ApJ...565..385U,Heger2002ApJ...567..532H}. This creation reduces the radiation pressure of the core, destabilizing it, and ultimately resulting in a thermonuclear explosion if the core is not too massive. If it is valid to extend the numerical modeling of massive stars up into this unobserved mass range, and if this mass range did exist at some time \citep{Chon2021MNRAS.508.4175C}, then we would expect to observe these explosions, via their supernovae lightcurves \citep{Moriya2022A&A...666A.157M}, their chemical enrichment \citep{Salvadori2019MNRAS.487.4261S}, or the absence of intermediate black holes which they prevent from forming \citep{Farmer2019ApJ...887...53F}. Both of the conditions for the existence of PISNe are thought to hold, and yet there is not convincing observational evidence for any of the consequences of PISNe explosions.

The merging intermediate mass black hole population, discovered via gravitational waves, could be consistent with PISNe preempting the formation of black holes with masses over about $60\; \msun$ \citep{Abbott2021ApJ...913L...7A}, but the discovery of a black hole with mass $80 \;\msun$ has raised questions about this interpretation. There are several suggestions for creating such a black hole in the paradigm of PISNe, from below the mass gap \citep{Samsing2021ApJ...923..126S}, from above \citep{Siegel2022ApJ...941..100S}, or directly from the stellar track \citep{Farmer2020ApJ...902L..36F,Umeda2020ApJ...905L..21U}. Because of the low prevalence of these types of systems, it is unlikely that current gravitational wave detectors will be able to resolve the existence or non existence of the mass gap. Thus, in this paper we focus on the lightcurves from these explosions. Although some low energy PISN are consistent with standard supernovae \citep{Kasen2011ApJ...734..102K}, most attempts to identify these events focus on super-luminous supernovae (SLSNe). Most SLSNe are thought to be powered by a central engine or interaction, but some are not consistent with these explanations, and remain candidates for PISNe.

\begin{figure*}
    \centering

    \includegraphics[width=1.0\linewidth]{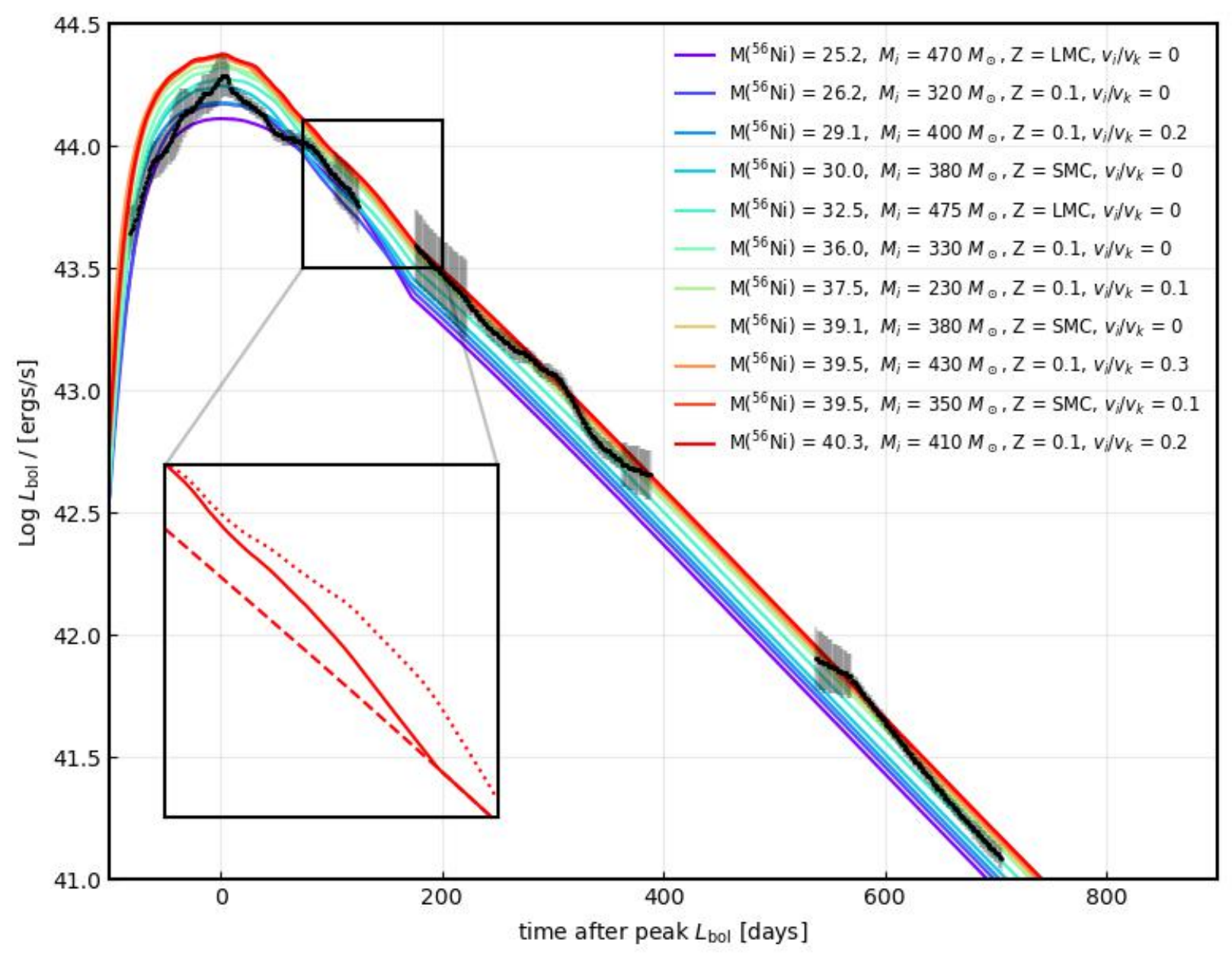}    
    \caption{Bolometric lightcurves of the models (colored by nickel mass) compared with observations of SN2018ibb. These lightcurves are a combination of the output of STELLA and the analytic expression for nickel decay (Eq. \ref{eq:dep}). The first part of the lightcurve shows the STELLA output, with the transition occurring between 100 and 300 days after the maximum luminosity as the ejecta becomes optically thin. For the most luminous model, the insert shows the luminosity from STELLA (dotted line) and the nickel deposition, (dashed line). The plotted bolometric luminosity transitions from the former to the latter within this time window. The observations of SN2018ibb \citep{Schulze2023arXiv230505796S} are shown in black with 1$\sigma$ errorbars in gray. Larger error bars generally denote less complete wavelength coverage. 
    }
    \label{fig:lbol}
\end{figure*}

\begin{figure*}
    \centering

    \includegraphics[width=1.0\linewidth]{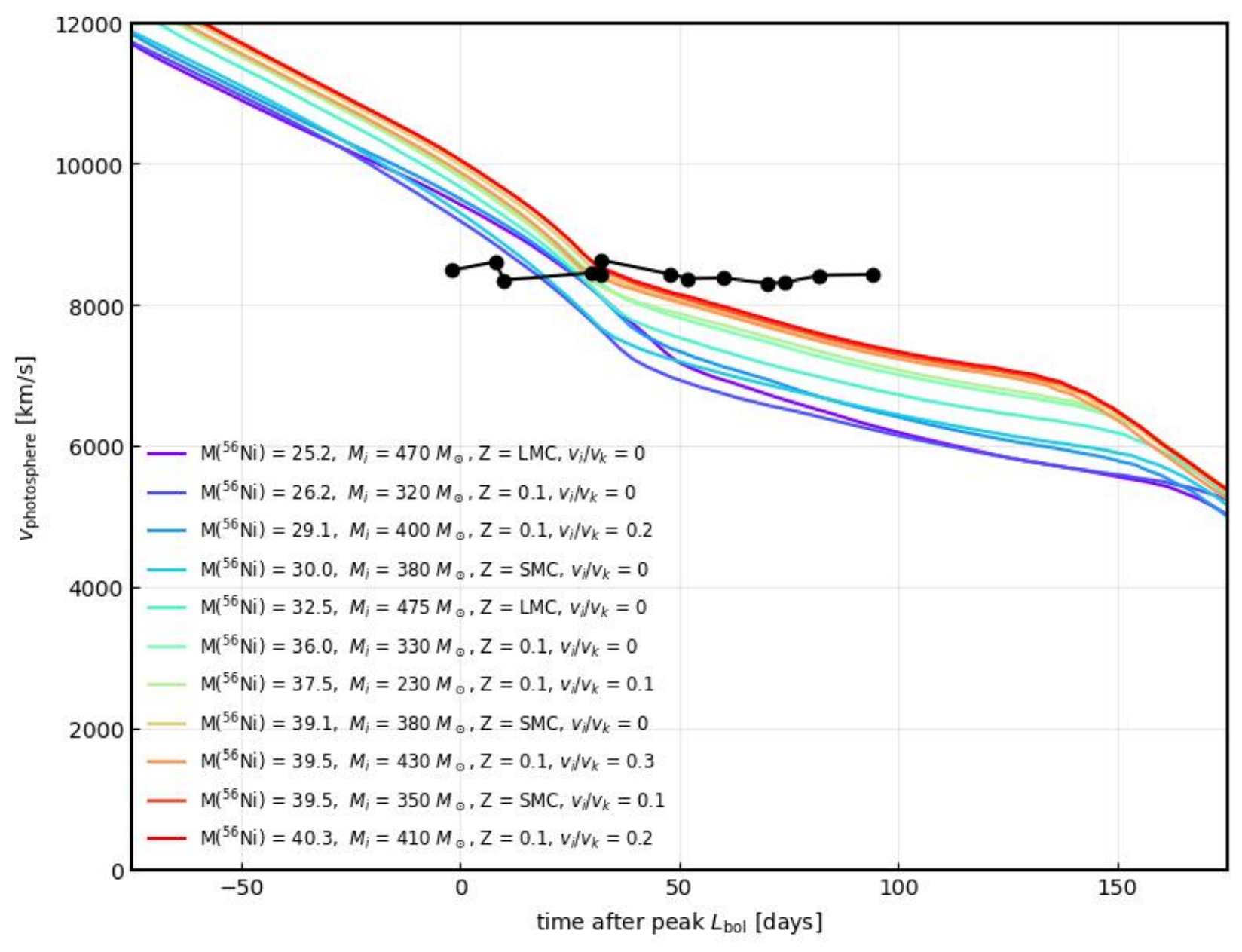}    
    \caption{Photospheric velocities of the models (colored by nickel mass) compared with observations of SN2018ibb \citep[black,][]{Schulze2023arXiv230505796S}. 
    }
    \label{fig:vel}
\end{figure*}

Recently, a promising PISN candidate was observed \citep[SN2018ibb][]{Schulze2023arXiv230505796S}. This SLSNe is characterized by two unique features: a lightcurve consistent with a nickel tail driven by $\sim 30 \;\msun$ of nickel, and evidence of circum-stellar medium (CSM) interaction. This paper reported multi-band observations (radio to x-ray) conducted from 70 days before the peak luminosity to 700 days after peak luminosity. In addition, polarimetry and optical/NIR spectra were taken at various times. Using these observations, the paper reconstructs a bolometric luminosity which is consistent with nickel decay (Fig. \ref{fig:lbol}). Crucially, the bolometric light curve is observed out to 706 days past the maximum luminosity. The decay of the luminosity in the later stages of the lightcurve (500 - 706 days) is exponential, allowing central engine models with power law decays to be ruled out. The exponential decay is not flawless, however, as a bolometric excess roughly 300 days after the peak luminosity is clearly visible. This excess is accompanied by enhanced line fluxes in [O II] and [O III], leading to the suggestion of the presence of an oxygen rich CSM. There is further evidence of CSM interaction in the blue side of the nebular spectra, taken at 637 days after the peak luminosity, which cannot be explained by a PISN alone. These two features, a supernova driven by substantial nickel decay and an oxygen rich CSM are both plausibly explained by the rotating, metal enriched PISNe models presented in \citet{Umeda2023arXiv230702692U}. In this letter, we discuss the observational properties of the relevant models in more detail.

\begin{figure*}
    \centering

    \includegraphics[width=1.0\linewidth]{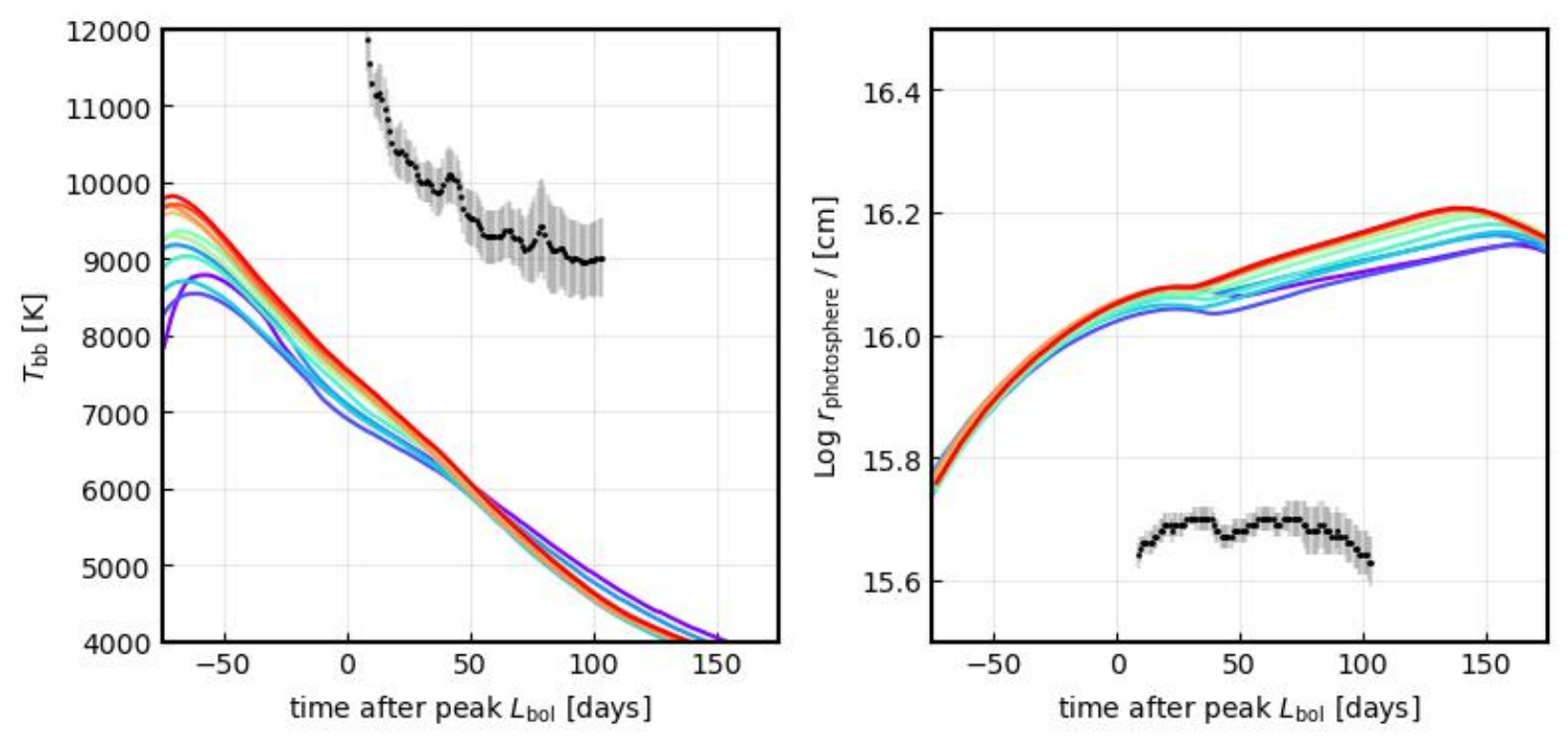}    
    \caption{Photospheric properties of the models (same colors as previous figures) compared with observations of SN2018ibb (grey shows 1$\sigma$ errorbars). Left panel --- black body temperature as a function of time. Right panel --- photosphere radius as a function of time. 
    }
    \label{fig:rt}
\end{figure*}

\section{\textbf{Methods}}
\label{methods}

The exploding PISN models in this letter are taken from \citet{Umeda2023arXiv230702692U}, in which we simulated the evolution and explosion of rotating metal enriched pair instability progenitors. \citet{Umeda2023arXiv230702692U} found a wide range of behavior, with nickel masses ranging from less than one solar mass up to more than forty, and a correspondingly large range in explosion energies. Here, we only consider models with nickel mass above twenty five, which are roughly consistent with the properties of SN2018ibb. 

For the lightcurve calculation, we use the publicly available version of STELLA \citep{Blinnikov1998ApJ...496..454B,Blinnikov2000ApJ...532.1132B,Blinnikov2006A&A...453..229B} which is included with MESA \citep[][]{Paxton2011ApJS..192....3P,Paxton2013ApJS..208....4P,Paxton2015ApJS..220...15P,Paxton2018ApJS..234...34P,Paxton2019ApJS..243...10P} version r23.05.1 to calculate the first few hundred days of the lightcurve. After the ejecta becomes optically thin, we revert to the analytic expression \citep{Jeffery1999astro.ph..7015J,Sharon2020MNRAS.496.4517S}. To do this, one may assume that nickel decay is the only source of luminosity, such that $L_{bol} = Q_{\rm dep}$ where $Q_{\rm dep}$ is the energy per unit time deposited in the ejecta, and is a combination of the energy deposited by gamma rays $Q_\gamma$ and the kinetic energy of the positrons $Q_{\rm pos}$. In practice, the deposition rate is modeled by the following expression, adopting a characteristic timescale ($t_0$) for the optically thin-to-thick transition for the gamma rays: 
\begin{equation}
    Q_{\rm dep} \approx Q_{_\gamma} (1-e^{-t_0^2/t^2}) + Q_{\rm pos} . 
    \label{eq:dep}
\end{equation}
As in \citet{Schulze2023arXiv230505796S}, we take $t_0 = 600$ days; this is consistent with the ejecta properties in the models for SN2018ibb, while we note that a range of the timescales are possible \citep{Maeda2003ApJ...593..931M}.

\begin{figure}
    \centering

    \includegraphics[width=1.0\linewidth]{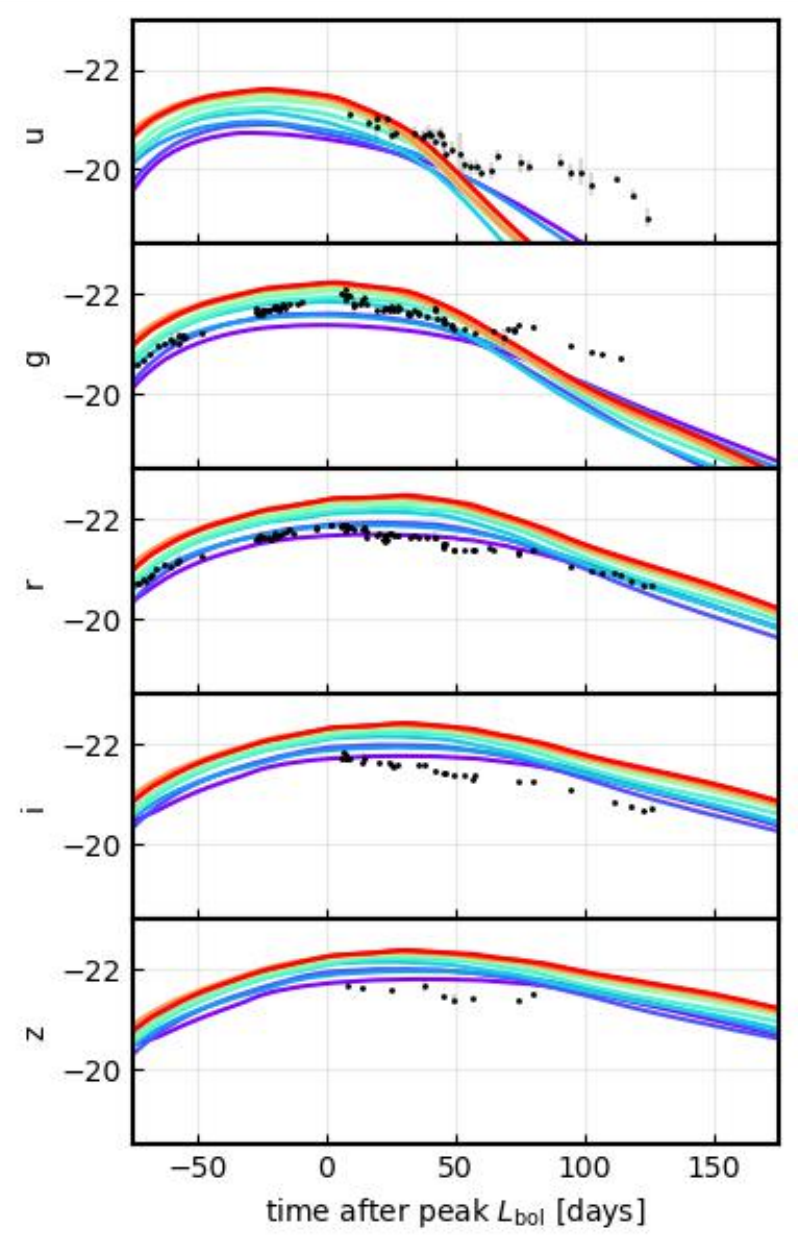}    
    \caption{AB magnitudes of the ugriz bands for the STELLA models (colored lines) compared to observations of SN2018ibb (black). The offset between the models and the observations as a function of wavelength is clearly visible, as reflected in the black body temperature offset (Fig. \ref{fig:rt})
    }
    \label{fig:Mab}
\end{figure}

We transport the final timestep of the hydrodynamical simulation from \citet{Umeda2023arXiv230702692U} (255 meshes) to STELLA (we use 509 meshes) after interpolation of log values of the hydrodynamical quantities and the elemental abundances. Note that we use the nickel abundance derived from the post-processing for consistency, though this value is lower than that found by the hydrodynamical simulations by a few percent due to a combination of bottlenecks in the nuclear network and the presence of isotopes around the iron peak in the larger network. The primary quantity of interest from STELLA is the bolometric luminosity, but we also utilize the photosphere location. We report the luminosity as a linear combination of the STELLA and analytic results, such that the relative weight is given by the fraction of nickel inside the photosphere at each timestep. This means that the luminosity will come directly from STELLA until about 100 days after the peak luminosity, and completely from the analytic expression after 200 days after the peak luminosity (Fig. \ref{fig:lbol}). In between these two times, we use a combination set by the photosphere mass coordinate as determined by STELLA. 

We also estimate the luminosity due to the interaction of the ejecta with a CSM density structure computed using the mass loss rate from the stellar evolution simulations and a variety of wind luminosities. Thus, for a given model the total CSM mass is fixed, but the density varies with the assumed wind speed. Even for extremely slow wind velocities ($\sim 10$ km/s), the CSM density did not become high enough to effect the early part of the lightcurve.
The estimation is done through the model of \citet{Maeda2022ApJ...927...25M} that includes the calculation of the energy conversion efficiency from the dissipated kinetic energy to radiation. Here, rather than directly using the results of the progenitor evolution and explosion calculations, we have adopted simplified ejecta structure (broken power-law) and CSM structure (a single power-law) that mimic the outcome of these simulations. 

To verify these estimates, we run hydrodynamics only simulations of the later phases of the explosion using the SuperNova Explosion Code \citep[SNEC, ][]{Morozova2015ascl.soft05033M}. These simulations include the exploding stellar models and the surrounding CSM, and allow us to estimate the change in kinetic energy of the ejecta as it sweeps up the CSM material. The rate of kinetic energy dissipation from realistic CSM densities in the SNEC models matches the dissipation in the broken power-law model described in the previous paragraph.

\section{\textbf{Results}}
\label{results}

First, we will discuss the bolometric lightcurve from STELLA in comparison with SN2018ibb (Fig. \ref{fig:lbol}). This comparison can be split into two sections, the first concerning the region around the peak luminosity which is taken from the STELLA simulations, and the second concerning the nickel decay phase which is computed analytically. In the first region, the luminosities of the PISNe models and SN2018ibb reach similar maximum values, but the morphology of the observed lightcurve is less broad. Though mixing of PISN ejecta is not thought to be efficient \citep{Chen2020ApJ...897..152C}, we have experimented with different nickel distributions in the STELLA simulations, while holding the total nickel mass constant. This mixing tends to broaden the peak lightcurve further and is thus not a suitable explanation for the difference. Another possible synthesis of the observation and the simulations would be if the true peak luminosity occurred earlier than is suggested by the black line (note that the grey errorbars show the one sigma values). If this were true, however, the shift in the peak luminosity would mean that the decay phase would be consistent with being powered by a larger amount of nickel, which in turn would suggest a larger peak luminosity. Another explanation proposed during analysis of the Type IIb SN2020acat is that radioactive heating creates nickel bubbles, the expansion of which can sharpen the peak lightcurve \citep{Ergon2022A&A...666A.104E,Ergon2024A&A...683A.241E} and the treatment of this phenomenon will differ between codes. Finally we note that STELLA has been shown to not fully reproduce the results of the Monte-Carlo radiation code SEDONA in some scenarios \citep{Tsang2020ApJ...898...29T}. Pivoting towards the decay phase, there are several wiggles in the lightcurve which are inconsistent with pure nickel decay in the optically thin stage. \citet{Schulze2023arXiv230505796S} noted that these wiggles may be due to deviations from blackbody emission though they also suggest that the wiggles at 300 and 600 days may originate from luminosity associated with the interaction of the explosion ejecta with CSM material. Of the two excesses, the first is the more robust as it is observed in nearly all of the available bands. However, the CSM structures required to explain both of the bumps simultaneously would have to be extremely inhomogeneous, suggesting that the origin of the CSM must have been discrete ejection events rather than due to mass loss on evolutionary timescales. Despite these caveats, however, the overall PISN bolometric lightcurves are remarkably consistent with SN2018ibb. The most consistent of our models is the LMC metallicity, $M_i = 475\;\msun$, $v_i/v_k=0$ model, which has a nickel mass of 32.5 $\msun$. We caution, however, that although the vertical errorbars towards the end of the bolometric lightcurve are small, the horizontal errorbars on the precise time of the peak luminosity are larger. Because we must rely on this peak time in order to calibrate the amount of nickel powering the late time emission, we feel that any of these models could be plausibly considered as lightcurves of SN2018ibb.

When it comes to the photosphere properties, however, reconciling these simulations with the observation becomes more challenging. Fig. \ref{fig:vel} shows the photosphere velocity. For the observation, this velocity is roughly constant at a value of just over 8000 km/s, whereas the simulations decrease from about 10000 km/s to 7000 km/s during this time period. While the values of the velocities are quite similar, the observation shows a constant velocity, while the simulations are decelerating. We note that the velocities from the simulations level off somewhat about 50 days after the peak luminosity as the photosphere begins to penetrate the higher opacity nickel region. This demonstrates that the disagreement between the observed velocity and the STELLA models matters more for what it implies about the velocity in the previous (unobserved) phase. If we extrapolate the constant velocity observation backwards in time, then there becomes a huge discrepancy in the velocity at 100 days before the peak luminosity. This is salient because the observed photosphere radius and black body temperature (Fig. \ref{fig:rt}) disagree more strongly with the STELLA models, and this can be traced back to the simulations having much higher velocities in the pre-luminosity-peak epoch. 

A simple formulation of this argument follows. If we multiply the rise time of 93.4 days by the observed velocity of 8000 km/s, then we find Log $r_{\rm photopshere} / \rm{cm} = 15.8$ which is much larger than the observed value of the photosphere (Fig. \ref{fig:rt}). This means that in order for the observations to be self consistent, either a) the velocity must have increased during this time period (contrary to expectation) or b) the observed photosphere is underestimated (Sec. \ref{discussion}). If we further require that the observations be consistent with PISN models, the situation is even more tenuous as velocities of up to $15000$ km/s are expected during the luminosity rise. To reinforce this point about the disagreement between the observations and the models regarding the photosphere properties, we show a direct comparison to five of the observed bands (ugriz, Fig. \ref{fig:Mab}). This figure shows that the observed supernova is bluer than the STELLA models, despite having a very similar bolometric luminosity (Fig. \ref{fig:lbol}). One possible explanation is that a contribution from the interaction luminosity could raise the UV flux, in an effect known as 'back shining' \citep{Dessart2022A&A...660L...9D}. However, this increase in the UV is thought to require a strong interaction, not present in the earlier part of this lightcurve, and it is unclear if any interacting luminosity can compete with the prodigious nickel emission. Due to the extreme scenario considered here, further study will be required to fully understand if this effect is relevant. There is also the possibility that a fraction of the gamma ray emission is stored during a series of degradation steps and released later on as UV emission \citep{Kozma1992ApJ...390..602K}, which could shift the observed temperature.

A general consideration is that for any PISN model, if the velocity is as low as the observation suggests, then the explosion cannot have produced $\sim 30 \;\msun$ of nickel, because there is a strong correlation between the explosion energy (which is then converted into kinetic energy) and the nickel mass. In other words, the derived photosphere radius is too small to be consistent with an \textit{energetic} PISN.

If we gloss over the inconsistencies outlined above, and instead posit that this observation is the result of a PISN, then what can we learn from the bolometric light curve? We would like to infer the initial properties of the progenitor, specifically the mass and rotation. Fig. \ref{fig:lbol} shows all models from \citet{Umeda2023arXiv230702692U} which produce more than 25 $\msun$ of nickel. \citet{Schulze2023arXiv230505796S} measured the metallicity of the host galaxy as $\zsun/4$, which falls in between the SMC ($\zsun/5$) and the LMC ($\zsun/3$). Thus, limiting ourselves to those two cases, we find that the initial masses of our stellar models consistent with this explosion are $350-475\;\msun$. The initial rotations of these stars include both the $v_i/v_k = 0$ and $v_i/v_k = 0.1$ cases. To get an idea of how common these explosions are, we compute that $8.7\times 10^{-6}$ of the mass in the Chabrier initial mass function (IMF) integrated from 0.01 $\msun$ to $1000$ $\msun$ falls into this explosion window \citep{Madau2014ARA&A..52..415M}. Very roughly, this means that there will need to be $4.0 \times 10^7\;\msun$ of stars formed before one star in the explosion window forms. This mass is the same order of magnitude as the current stellar mass of the host galaxy of SN2018ibb \citep[Log $M/\msun = 7.6^{+0.19}_{-0.22}$,][]{Schulze2023arXiv230505796S} and at this galaxy's current star formation rate ($0.02^{+0.04}_{-0.01}\;\msun$/yr), it would take two billion years to produce that much stellar mass. On the other hand, we note that stars with initial masses approaching the explosion window have been observed in the LMC \citep{Doran2013A&A...558A.134D} and an analysis based on a power law IMF may underestimate the prevalence of such massive stars.

\section{\textbf{Discussion}}
\label{discussion}

We have run STELLA simulations of energetic pair instability supernovae originating from metal enriched rotating progenitors in view of a comparison to SN2018ibb. The stellar progenitors are thought to be metal enriched because of the metallicity of the host galaxy, and because of hints at the presence of an oxygen rich CSM surrounding the star. Rotation and the presence of the metals in the progenitor both contribute to stripping the envelope of the star, producing this CSM. In this paper, we focus on simulations of energetic explosions which produce more than 25 $\msun$ of nickel. This is because SN2018ibb appears to be powered by nickel decay more massive than this value. We found that the STELLA models, in concert with analytic expressions for the luminosity due to nickel decay in the later phases, can reproduce the overall shape of the bolometric light, but not specific features such as the narrowness of the peak luminosity or the excesses during the optically thin phase. Our analysis of the luminosity contributed by the interaction of the shockwave with the evolutionary CSM can contribute to the overall luminosity and might be visible in the spectrum, as has been observed, but will not be large enough to cause the bumps in the bolometric light curve. 

The STELLA models cannot, however, reproduce other observational aspects of this supernova. In particular, the deceleration of the photosphere found in the models does not match the observational behavor, and, as a kind of corollary, the radius and temperature of the photosphere are vastly different between those inferred by the observations and predicted by the models. Because of the nearly monotonic correlation between nickel mass and kinetic energy of the ejecta in these thermonuclear supernovae, a supernova with this large nickel mass and low velocity is hard to reconcile with any PISN model. This disagreement, however, may be simply due to our incomplete understanding of line blanketing near the photosphere, which may push the spectrum up or down in wavelength \citep[c.f. SNIa, ][]{Sauer2008MNRAS.391.1605S,Walker2012MNRAS.427..103W}.

If this event was not a PISN, then what was it? As seen in Table 8 of \citet{Schulze2023arXiv230505796S}, the lightcurve of this supernova can be readily explained as an interacting supernova, but this requires the fine-tuning of the CSM density such that the lightcurve mimics nickel decay, and this explanation also seems to be incompatible with the nebular spectra (Fig. 27). Also in this table, the scenario of a magnetar combined with 32 $\msun$ of nickel is considered. Such a model would explain both the narrowness of the peak luminosity and the large nickel decay tail, but the model parameters were rejected for being unphysical (e.g. $M_{\rm ejected} = 54 - 141 \;\msun$ are not typical values associated with neutron star formation). \citet{Schulze2023arXiv230505796S} do not consider fallback accretion plus nickel decay, and although at least one mechanism has been proposed for this scenario \citep{Siegel2022ApJ...941..100S}, it would be even rarer than PISNe. Central engine models have been suggested to have the constant velocity behavior seen in Fig. \ref{fig:vel} \citep{Kasen2010ApJ...717..245K}, but this one dimensional result has been shown to break down in higher dimensions due to the effects of Rayleigh–Taylor instabilities \citep{Suzuki2019ApJ...880..150S,Suzuki2021ApJ...908..217S}.

In summary, we have presented forward modeling of metal enriched pair instability supernovae culminating in lightcurves for comparison to SN2018ibb. We find that the STELLA models can reproduce the observed bolometric luminosity and velocity of this supernova, but there are outstanding issues related to the velocity evolution, rise time, and photosphere properties of this supernova which require further study.

\section*{Data Availability}

The data underlying this article will be shared on reasonable request to the corresponding author.

\bigskip \bigskip

\section*{Acknowledgements}

This study was supported in part by JSPS KAKENHI Grant Numbers JP20H00174, JP23K20864, JP24K00682, JP24H01810, and JP24K00682.


\bibliography{bib}{}
\bibliographystyle{aasjournal}



\end{document}